\documentclass[prl,superscriptaddress,longbibliography,footinbib,twocolumn,preprintnumbers,amsmath,amssymb]{revtex4-2}

\usepackage{xr}
\makeatletter
\newcommand*{\addFileDependency}[1]{
	\typeout{(#1)}
	\@addtofilelist{#1}
	\IfFileExists{#1}{}{\typeout{No file #1.}}
}
\makeatother

\newcommand*{\myexternaldocument}[1]{%
	\externaldocument{#1}%
	\addFileDependency{#1.tex}%
	\addFileDependency{#1.aux}%
}

\myexternaldocument{supp}

\usepackage{amssymb}
\usepackage{amsmath}

\usepackage{color}
\usepackage[pdftex]{graphicx}

\usepackage{empheq}

\usepackage{mathtools}

\usepackage{makecell}

\usepackage{enumerate}
\usepackage{mathtools}
\usepackage{stmaryrd}

\usepackage{enumitem}

\usepackage{textcomp}
\usepackage{gensymb}

\newcommand{\rr}[0]{\boldsymbol{r}}

\newcommand{\kB}[0]{k_{\mathrm{B}}}

\usepackage{lipsum}  
\newcommand{\cC}{\mathcal{C}}

\renewcommand{\ss}{\boldsymbol{s}}

\definecolor{darkblue}{rgb}{0,0,0.6}
\definecolor{darkred}{rgb}{0.6,0,0}
\usepackage[colorlinks=true,urlcolor=darkblue,citecolor=darkblue,linkcolor=darkred]{hyperref}

\begin{document}

\title{Chemically active droplets in crowded environments}

\author{Jacques Fries}
\author{Roxanne Berthin}
\affiliation{Sorbonne Universit\'e, CNRS, Physico-Chimie des \'Electrolytes et Nanosyst\`emes Interfaciaux (PHENIX), 4 Place Jussieu, 75005 Paris, France}

\author{Chengjie Luo}
\affiliation{Max Planck Institute for Dynamics and Self-Organization, Am Fa{\ss}berg 17, 37077 G\"ottingen, Germany}

\author{Marie Jardat}
\affiliation{Sorbonne Universit\'e, CNRS, Physico-Chimie des \'Electrolytes et Nanosyst\`emes Interfaciaux (PHENIX), 4 Place Jussieu, 75005 Paris, France}

\author{David Zwicker}
\affiliation{Max Planck Institute for Dynamics and Self-Organization, Am Fa{\ss}berg 17, 37077 G\"ottingen, Germany}

\author{Vincent Dahirel}
\affiliation{Sorbonne Universit\'e, CNRS, Physico-Chimie des \'Electrolytes et Nanosyst\`emes Interfaciaux (PHENIX), 4 Place Jussieu, 75005 Paris, France}

\author{Pierre Illien}
\thanks{pierre.illien@sorbonne-universite.fr}
\affiliation{Sorbonne Universit\'e, CNRS, Physico-Chimie des \'Electrolytes et Nanosyst\`emes Interfaciaux (PHENIX), 4 Place Jussieu, 75005 Paris, France}

\begin{abstract}

Biomolecular condensates are essential for cellular organization and result from phase separation in systems far from thermodynamic equilibrium. Among various models, chemically active droplets play a significant role, consisting of proteins that switch between attractive and repulsive states via nonequilibrium chemical reactions. While field-based simulations have provided insights into their behavior, these coarse-grained approaches fail to capture molecular-scale effects, particularly in crowded cellular environments. Macromolecular crowding, a key feature of intracellular organization, strongly influences molecular transport within condensates, yet its quantitative impact remains underexplored. This study investigates the interplay between chemically active droplets and crowders by using particle-based models, that provide molecular insight, and a field-based model, that complements this picture. Surprisingly, crowding reduces droplet size while expanding the overall dense phase volume, challenging equilibrium-based expectations. This effect arises from the interplay between depletion interactions, diffusion hindrance, and nonequilibrium particle fluxes. Our findings provide a step towards a more comprehensive understanding of chemically active droplets in complex, realistic cellular environments.
\end{abstract}

\date{\today}

\maketitle

\emph{Introduction.---}  Biomolecular condensates, which play a key role in the spatial organization of the cytoplasm and of the nucleus of biological cells, have become a central topic in biophysics \cite{Hyman2014,Gouveia2022,Shin2017,Alberti2019,Alberti2019a}. These microscopic or nanoscopic assemblies of proteins typically result from a local phase separation phenomenon, that occurs in a system maintained very far from thermodynamic equilibrium: this makes their modeling particularly challenging.

Among the different models of biomolecular condensates, chemically active droplets now play an important role \cite{Weber2019}. They are assumed to be made of proteins that coexist under two forms: in a form called droplet material, the proteins attract each other through short-range interactions, and in another form, they simply repel each other and do not take part in the formation of droplets. The transitions between these two forms occur via nonequilibrium  chemical reactions \cite{Zwicker2015,Zwicker2017,Wurtz2018,Kirschbaum2021}. Such chemically active droplets have been studied through field-based simulations \cite{Glotzer1994,Christensen1996,Zwicker2022}, where the interactions between the different species is encoded in a free energy functional, and which have recently been extended to more complex descriptions, that account for electrostatic interactions \cite{Luo2025}, the effect of confinement \cite{Rossetto2024}, or the role of enzyme molecules \cite{Cotton2022}.

Nevertheless, this coarse-grained approach cannot predict accurately the structure and dynamics of biomolecular condensates in situations where microscopic details matter the most \cite{Dignon2020}. For instance, a central aspect of intracellular organization is macromolecular crowding: metabolic processes are known to take place in strongly dense environments \cite{Zimmerman1993,Ellis2001,Ellis2003,Hall2003,Rivas2016}, and biomolecular condensates thus operate under conditions where the transport of droplet material is strongly affected by the presence of crowders \cite{Delarue2018,Andre2020,Spruijt2023}. Even though the effect of macromolecular crowding on  the phase separation of {\it in vitro} polymers at equilibrium is well known \cite{Alfano2024}, its role in the context of biomolecular condensates has not been studied in minimal physical models yet.

In this Letter, we study the interplay between chemically active droplets and crowders through two numerical approaches:  a particle-based description, where the proteins and crowders are represented explicitly as interacting Brownian particles \cite{Berthin2025,Fries2024}, and a field-based description, where the crowders are represented as an additional density field \cite{Zwicker2022}. In both approaches, we study the effect of crowding on the structure and dynamics of active emulsions. In agreement with the behavior of equilibrium systems, the presence of crowders increases the overall volume of the dense phase. This reflects the influence of thermodynamic driving forces, such as depletion interactions. However, this larger volume is surprisingly split into smaller droplets, against the intuition that crowding favors condensation. We show that different crowding-induced mechanisms, such as diffusion hindering or chemical reaction acceleration, contribute to this observation. This work is a first step towards a quantitative description of chemically active droplets in complex environments, going beyond idealized descriptions.

\emph{Particle-based simulations.---}  We first simulate a three-dimensional suspension of interacting particles, obeying Brownian (i.e. overdamped Langevin) dynamics, which are integrated with a standard algorithm \cite{Allen1987,Thompson2022} (see Section I of the Supplemental Material \cite{SM} for details on numerical simulations). The suspension is made of three species, denoted by $A$, $B$ and $C$. The $B$ particles interact via an attractive Lennard-Jones potential and are therefore the droplet material \cite{Berthin2025}. The interaction parameter $\varepsilon$ of this potential is chosen in such a way that the $B$ particles show liquid-gas coexistence at the considered density and at equilibrium.  All the other interactions between particles are repulsive (we use the Weeks-Chandler-Andersen potential \cite{Weeks1971} to model their interactions).

$B$ particles may convert into $A$ particles, and vice versa, and the corresponding conversions occur via two distinct pathways: a passive pathway, in which detailed balance is fulfilled, and an active pathway, in which detailed balance is broken~\cite{Berthin2025}. Let us denote by $\mathcal{C}$ and $\mathcal{C}'$ two configurations which only differ by the species of one particle, and by $k^{\text{p}}_{\cC,\cC'}$ (resp. $k^{\text{a}}_{\cC,\cC'}$) the rate at which the conversion from $\mathcal{C}$ to $\mathcal{C}'$ occurs through the passive (resp. active) pathway. On one hand, the detailed balance conditions imposes ${ k^{\text{p}}_{\cC',\cC}}/{k^{\text{p}}_{\cC,\cC'}} = \exp\{-\beta[E(\cC)-E(\cC')]\}$, where $\beta=(\kB T)^{-1}$, and where $E(\cC)$ is the energy of configuration $\cC$. On the other hand, the active reaction rates fulfill ${ k^{\text{a}}_{\cC',\cC}}/{k^{\text{a}}_{\cC,\cC'}} = \exp\{-\beta[E(\cC)-E(\cC')+\kappa_{\cC',\cC}\Delta \mu]\}$ where $\Delta \mu$ is a chemical drive fixed to $4\kB T$, and $\kappa_{\cC',\cC}=1$ if the transition from $\cC'$ to $\cC$ implies the formation of a $B$ particles and $-1$ otherwise. The prefactors that remain to be determined are chosen in such a way that passive (resp. active) reactions dominate in the dilute (resp. dense) phase \cite{Berthin2025}. Finally, the particles of species $C$ do not take part in the reactions, and  play the role of crowders. While the volume fractions of $A$ and $B$ is fixed ($\phi_A+\phi_B=0.025$), the volume fraction of the crowders is tuned in such a way that the total volume fraction $\phi=\phi_A+\phi_B+\phi_C$ varies from $0.05$ to $0.425$.

\begin{figure}
    \centering
    \includegraphics[width=\columnwidth]{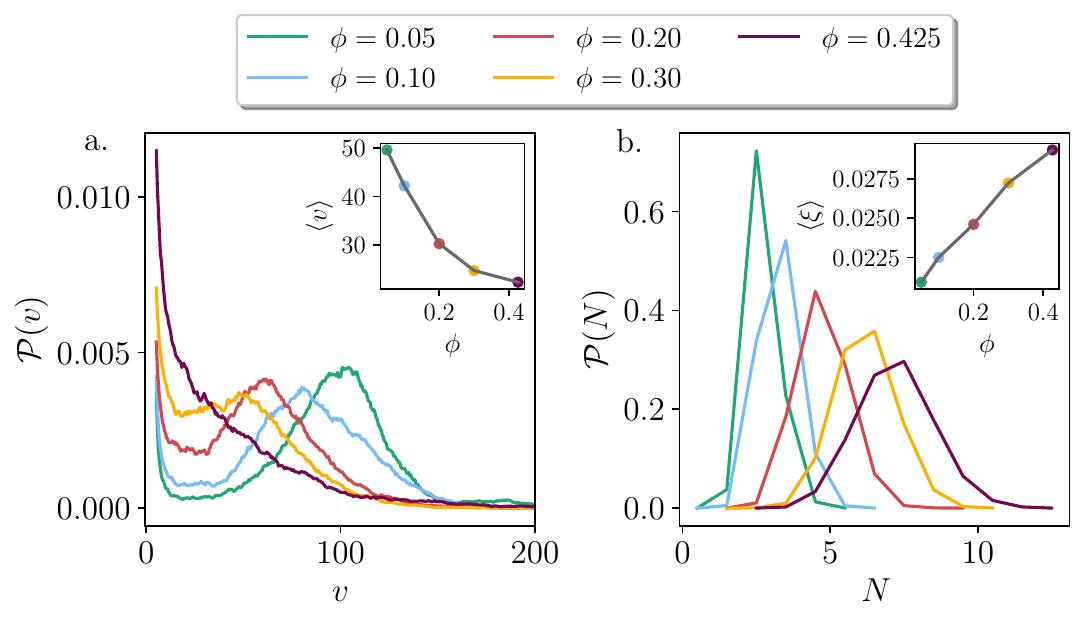}
    \caption{
    (a) Probability distribution of the droplet volume in the stationary state, for various values of the total volume fraction $\phi$. The inset shows the average volume of a droplet $\langle v \rangle$ as a function of the total volume fraction $\phi$.
    (b) Probability distribution of the number of droplets in the simulation box for various values of the total volume fraction $\phi$. The inset shows the average volume fraction of the dense phase $\langle \xi \rangle$ as a function of the total volume fraction $\phi$ (the color of each symbol corresponds to the same $\phi$ values that the ones on histograms shown on panels (a) and (b)).
    }
    \label{fig1}
\end{figure}

We previously showed that, in these simulations, the interplay between pair interactions and nonequilibrium reactions (tuned by the chemical drive $\Delta \mu$) leads to the formation of stable active droplets, which are continuously fed from the outside and destroyed from the inside. Within the framework of active emulsions \cite{Weber2019,Zwicker2022}, the droplets are `externally maintained'. In order to study the effect of crowding on the structure of the resulting active emulsion, the evolution of the system is simulated until it reaches a nonequilibrium steady state. We then compute the volume distribution of the droplets, for different values of the overall volume fraction $\phi$ [Fig.~\ref{fig1}(a)].  Strikingly, we observe that the average droplet volume (shown in the inset of Fig.~\ref{fig1}(a)) is a decreasing function of the volume fraction $\phi$. In order to check whether this diminution of the droplet volume is associated with an overall shrinking of the dense phase, we also plot on the inset of Fig.~\ref{fig1}(b) the  average volume fraction of the dense phase $\langle \xi \rangle$, defined as the total volume occupied by droplets divided by the volume of the simulation box. We see that $\langle \xi \rangle$ increases with crowding, indicating that, although the volume of the droplets decreases, they must be more numerous.  This is confirmed by the main plot on Fig.~\ref{fig1}(b), that shows the distribution of the number of droplets for different densities of crowders.

These findings go against what is typically expected from an equilibrium perspective. Indeed, at equilibrium, increasing crowding in such a simple setting is expected to encourage the condensation of larger and larger droplets \cite{Alfano2024}. Moreover, we show in Section II of the SM \cite{SM} that, in the absence of any reactions, the critical interaction energy $\varepsilon$ above which the $B$ particles form a single cluster decreases with the overall density $\phi$. We also show that the volume of the single droplet that is formed through successive coarsening events up to the stationary state is an increasing function of crowding.

We now aim at understanding the mechanisms that govern these counter-intuitive effects. Naively, one would expect that increasing the density of crowders has the following consequences: (i) the attraction between the $B$ particles should be stronger, because of the depletion induced by the crowders \cite{Oosawa1954, Asakura1958, Tuinier2003,  Binder2014,Lekkerkerker2024} -- note that we will use the term `depletion' even though $A$, $B$ and $C$ have the same size; (ii) the mobility of the particles is expected to be smaller when $\phi$ increases \cite{Sokolov2012,Hofling2013,Benichou2018}. Note that, in an equilibrium situation (i.e. without reactions, or with $\Delta\mu=0$), since the stationary state is independent of the dynamics, only the effect of depletion actually matters.  In what follows, we study separately these effects, in order to understand how they affect the size and number of chemically active droplets that coexist at stationary state.

\begin{figure}
    \centering
    \includegraphics[width=0.8\columnwidth]{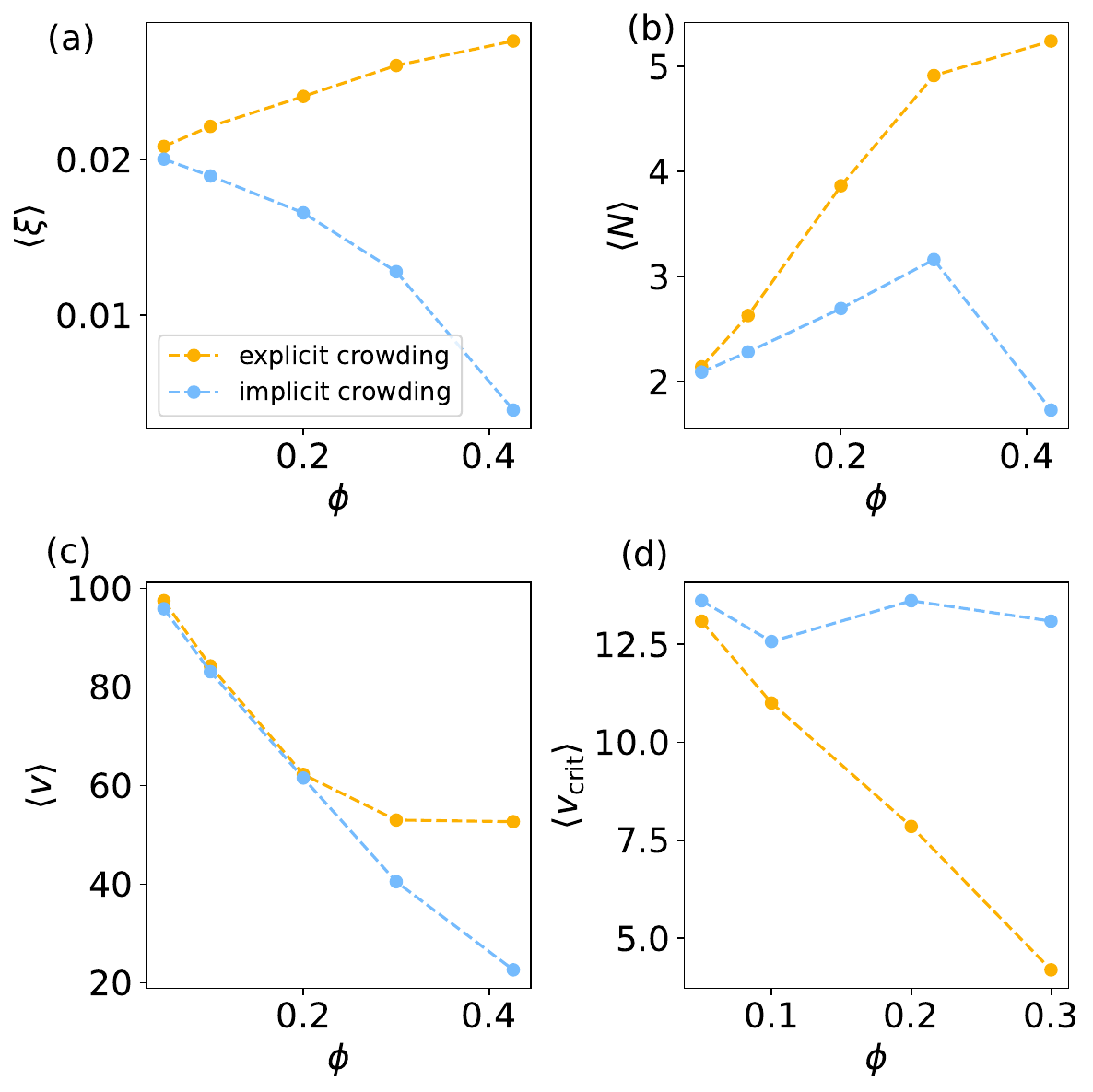}
    \caption{(a) Average volume fraction of the dense phase $\langle \xi \rangle$, defined as the total volume occupied by droplets divided by the volume of the simulation box, (b) average number of droplets $\langle N \rangle$, (c) average volume of the droplets $\langle v \rangle$, and (d) average critical nucleation volume of the droplets $\langle v_\text{crit}\rangle$, as a function of the overall volume fraction $\phi$, both in simulations where the crowders are represented explicitly (orange symbols) and implicitly (blue symbols). }
    \label{fig2}
\end{figure}

We start by investigating the effect of crowding on the mobility of the particles, and therefore on the diffusive fluxes that continuously feed and destroy the droplets. To this end, we perform simulations where the crowders are not represented explicitly, but are accounted for implicitly, through the bare diffusion coefficients of the $A$ and $B$ particles, that become functions of the volume fraction of crowders $\phi_C$. The diffusion coefficient for hard spheres in a three-dimensional system can be approximated as: $ D(\phi_C) \simeq D_0(1-2\phi_C)$ (note that this actually exact up to order $\mathcal{O}(\phi_C^2)$~\cite{Ackerson1982,Hanna1982, Lekkerkerker1984}), where $D_0$ is the diffusion coefficient of the particles in an infinitely dilute solution. This is the choice we make for the diffusion coefficients of $A$ and $B$ particles in these simulations, which will be called `implicit', as opposed to the previous ones, that will be called `explicit'. Importantly, the implicit simulations account for the fact that the diffusion coefficient of the particles decreases with crowding, but not for the fact that their interactions are effectively more attractive due to depletion interactions, which allows to decouple the two effects.

In this set of simulations, we measure the average volume fraction of the dense phase $\langle \xi\rangle $, the average number of droplets $\langle N \rangle$ and their average volume $\langle v \rangle$. These observables are shown respectively on Figs. \ref{fig2}(a), (b) and (c), as a function of the total volume fraction $\phi$. We observe that, in the implicit simulations, the average volume fraction of the dense phase $\langle \xi \rangle$ decreases [Fig.~\ref{fig2}(a)], whereas the number of droplets increases to a much lesser extent, up to the point where they become unstable [Fig.~\ref{fig2}(b)]: this means that the average volume of the droplets decreases, as seen on Fig.~\ref{fig2}(c). Therefore, although the implicit simulations correctly predict the diminution of the droplet size with increasing crowding, they do not account for the fact that the volume of the dense phase increases.

Finally, we also plot the critical nucleation volume as a function of the total volume fraction, for both implicit and explicit simulations (see Fig.~\ref{fig2}(d), where $\langle v_\text{crit} \rangle$ is measured from the phase portraits of the droplet volume \cite{Berthin2025}, which are shown in Section III  of the SM \cite{SM}). In implicit simulations, the nucleation volume is essentially independent of the total volume fraction. On the contrary, it strongly decreases with crowding in explicit simulations. In other words, in explicit simulations, small droplets are more likely to be stable against fluctuations, and it is therefore easier and easier to nucleate droplets as crowding increases: hence there are more droplets in explicit simulations than in implicit ones.

In summary, by comparing the two sets of particle-based simulations, we conclude that implicit simulations cannot account for the following observations from the explicit simulations: (i) crowding increases the total volume of the dense phase; and (ii) crowding reduces the critical nucleation volume. Both effects are then related to either the effect of depletion, or to the way active reactions, that are typically density-dependent, are implemented.

\emph{Field-based simulations.---}  To get further insight into the roles of depletion and active reactions, we also perform field-based simulations of active emulsions, in the spirit of Refs. \cite{Kirschbaum2021,Zwicker2022,Zwicker2025}. In these simulations, the system comprises the three species $A$, $B$, and $C$, just like in particle-based simulations, and a fourth species $S$, which plays the role of solvent.
The state of the system is then described by four density fields $\phi_i(\rr,t)$, which obey $\sum_{i\in \{A,B,C,S\}}\phi_i (\rr,t)=1$, so we do not need to describe $\phi_S$ explicitly. The fields for $A$, $B$, $C$ obey the reaction-diffusion equations $\partial_t \phi_i=\nabla \cdot (\lambda_i \phi_i \nabla \mu_i) + R_i$, where $\lambda_i$ denotes the diffusivity of species $i$ and the net reaction fluxes $R_i$ are
\begin{multline}
\label{eq:R}
    R_B =-R_A=2k_p(1-\phi_A-\phi_B)
        \sinh\Bigl(\frac{\beta}{2}(\mu_A-\mu_B)\Bigr)
\\
    +2k_a(\phi_A+\phi_B)\sinh\Bigl(\frac{\beta}{2}(\mu_A-\mu_B-\Delta\mu)\Bigr),
\end{multline}
whereas $R_C=0$. The two lines in Eq.~\eqref{eq:R} respectively describe passive and active contributions with rates $k_p$ and $k_a$, consistent with the particle-based setup.
We show in Section IV of the SM \cite{SM} that the results that follow are robust to changes of the functional dependence of the prefactors to the $\sinh$ functions, provided that the passive (resp. active) reaction dominate in the dilute (resp. dense) phase.
Chemical reactions are driven by the exchange chemical potentials $\mu_i = v\beta \delta F/\delta \phi_i$ following from free energy functional~$F$.
For simplicity, we consider a  Flory-Huggins free energy,
\begin{multline}
\label{eq:F}
    F=\frac{\kB T}{v} \int \bigg[
    w_B\phi_B + \sum_{i,j\in\{A,B,C,S\}}\chi_{ij}\phi_i\phi_j
\\
    +\sum_{i\in \{A,B,C,S\}}\phi_i\ln \phi_i + \frac{\kappa^2}{2}  |\nabla \phi_B|^2\bigg]\mathrm{d}\rr,
\end{multline} 
where the integral is over the entire system and $v$ denotes the molecular volume, which is the same for all species.
The first two terms in the square bracket describe the physical interactions among the species, the third term captures translational entropies of all four species, and the last term limits the width of interfaces between coexisting phases to roughly $\kappa$ in strongly interacting systems.
The interaction parameters $\chi_{ij}$ are defined as  $\chi_{ij}=e_{ij}-(e_{ii}+e_{jj})/2$, where $e_{ij}$ characterizes the interaction strength between species $i$ and $j$.
Specifically, we set $e_{BB} = E_\text{att}= -5\kB T$ to model the attractive interaction between $B$ particles, and $e_{ij} = E_\text{rep}= \kB T$ for all other pairs with $i,j \in \{A,B,C\}$, placing the system in a regime where macrophase separation occurs at equilibrium in the absence of reactions \cite{SM}.
We set the internal energy of species $B$ to $w_B=-4\kB T$ to favor the passive conversion of $A$ to $B$.
Conversely, the active reaction in Eq.~\eqref{eq:R} is driven by a chemical potential difference $\Delta\mu=4\kB T$, which drives the conversion of $B$ to $A$.
We simulate these equations in only one dimension since size-control of chemically active droplets is similar in all dimensions~\cite{Zwicker2025}.

\begin{figure}
    \centering
    \includegraphics[width=0.9\columnwidth]{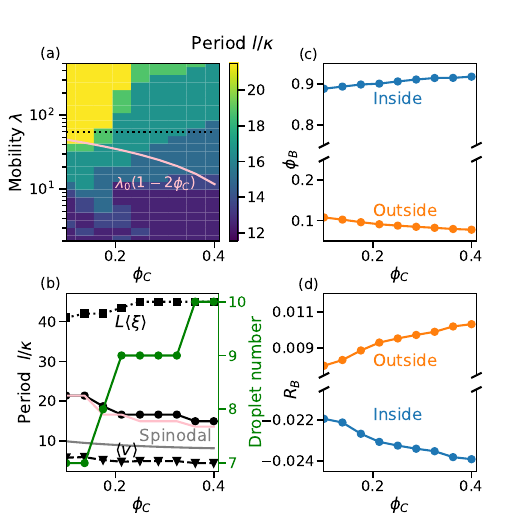}
      \caption{
      (a) Pattern period~$l$ as a function of the crowder density $\phi_C$ and the mobility $\lambda$ of all species. The functional form $\lambda(\phi_C)$ used in the implicit particle-based simulations is shown as the pink solid line.
      (b) Pattern period~$l$ (black disks), mean droplet volume $\langle v \rangle$ (triangles), mean fraction of the system taken up by the dense phase $\langle \xi \rangle$ (squares), and droplet number $L/l$ (green disks) as a function of $\phi_C$ for fixed mobility $\lambda=\lambda_0=58.6$ (corresponding to the black dotted line in panel (a)). The pattern period $l$ corresponding to the  mobility $\lambda_0(1-2\phi_C)$ is shown in pink, aligning with the pink curve in panel (a), whereas the grey line represents the most unstable length scale predicted from linear stability analysis.
      (c) Minimal and maximal values of $\phi_B$ as a function of $\phi_C$, corresponding to typical fractions outside and inside droplets, respectively.
      (d) Net reaction flux $R_B$ inside (blue) and outside (orange) of droplets as a function of $\phi_C$.
      (a--d) Additional parameters are $w_B=-4\kB T$, $\phi_A+\phi_B=0.4$, $k_a=k_p=0.01$, $\Delta\mu=4\kB T$, and linear system size $L=150\, \kappa$.}
    \label{fig4}
\end{figure}

We first analyze the typical droplet size emerging from the field-based simulations by  measuring the period $l$ of the solutions shown in Section IV of the SM \cite{SM}. Fig. \ref{fig4}(a) shows that $l$ typically decreases with increasing crowder density $\phi_C$, for all tested mobilities~$\lambda$ of the species $A$, $B$, and $C$.
While this trend is consistent with the particle-based simulations, crowders in principle also slow down the $A$ and $B$ molecules.
To capture this effect, we added the pink solid line to Fig.~\ref{fig4}(a), which captures the typical dependency of $\lambda$ on $\phi_C$.
The period length along this line (pink curve in Fig. \ref{fig4}(b)) shows a similar trend to the droplet size shown in Fig. \ref{fig2}(c), indicating that crowding generally decrease the size of chemically active droplets.

Our field-theoretic simulations allow disentangling the effect of depletion from the slow-down due to crowding.
Fig.~\ref{fig4}(b) shows that the period~$l$ and the average droplet size $\langle v\rangle$ decrease with increasing $\phi_C$ for fixed mobility $\lambda$, so we change the depletion effect without affecting the diffusive kinetics.
Concomitantly, the droplet number (green line) and the mean fraction $\langle \xi \rangle$ of the system taken up by the dense phase increase, consistent with the explicit crowding simulations shown in Fig.~\ref{fig2}. Moreover, Fig.~\ref{fig4}(c) shows that droplets get denser while the dilute phase thins out, indicating that phase separation becomes stronger with increasing $\phi_C$. We hypothesize that these compositional changes strongly affect the reaction fluxes~$R_i$ given by Eq.~\eqref{eq:R}, possibly explaining the effect on droplet size.
Indeed, Fig.~\ref{fig4}(d) shows that $|R_B|$ increase with larger $\phi_C$, indicating that droplet material $B$ is destroyed more quickly inside the droplets, while its creation in the dilute phase is also accelerated.

\emph{Discussion.---} A simple analytical argument can be invoked to rationalize some of our findings. The classical Lifshitz-Slyozov-Wagner theory of Ostwald ripening can be modified to account for active reactions \cite{Zwicker2015}. Within this framework, in the limit where droplets are sufficiently large to neglect surface tension effects, the typical radius $\mathcal{R}(t)$ of an isolated droplet evolves as (up to numerical prefactors): $\frac{\mathrm d\mathcal{R}}{\mathrm dt} = \frac{\mathcal{D}_B \phi_B^0}{\mathcal{R}}-\mathcal{K}_{BA} \mathcal{R}$, where $\mathcal{D}_B$ is the typical diffusivity of $B$ particles, $\phi_B^0$ the typical density of $B$ particles in the dilute phase, and $\mathcal{K}_{BA} $ is the typical rate at which the $B\to A$ reactions take place in the droplets. One finds that the typical droplet size in the stationary state can be estimated as $R\sim \sqrt{ \mathcal{D}_B \phi_B^0/\mathcal{K}_{BA}}$.
This simple scaling analysis shows how the main consequences of increased crowding (the reduction of the typical  diffusivity $\mathcal{D}_B$, the diminution of the typical density in the dilute phase $\phi_B^0$, and the enhancement of the reaction rates $\mathcal{K}_{BA}$) all result in an effective reduction of droplet size.

Our particle-based and field-based simulations take this simple analysis much further, by accounting finely for the effect of depletion interactions and for presence of multiple droplets. The three sets of numerical simulations presented in the manuscript (particle-based simulations with both explicit and implicit crowders on one hand, and field-based simulations on the other hand) allowed us to identify the role played by crowding on the structure of active emulsions, together with the underlying mechanisms. We generally observe that, when the volume fraction of crowders increases, the total volume occupied by droplets increases, but that the emulsion is composed of a larger number of smaller droplets. We showed that the reduction of the diffusive fluxes that results from crowding cannot be the only responsible for these observations, as it does not account for neither the overall increase of the volume of the dense phase, nor the facilitated nucleation of droplets. These two effects essentially originate from the depletion interactions induced by the presence of crowders, as confirmed by field-based simulations.

\emph{Acknowledgements.---} We gratefully acknowledge funding from the Max Planck Society and the European Research Council (ERC, EmulSim, 101044662).


%

\end{document}


\beginsupplement

\onecolumngrid

\begin{center}
	
	{\bf \large Chemically active droplets in crowded environments}

	$ \ $

	\textbf{\textit{\large Supplemental Material}}

	$ \ $

	Jacques Fries,$^1$ Roxanne Berthin,$^1$ Chengjie Luo,$^2$ Marie Jardat,$^1$ David Zwicker,$^2$ Vincent Dahirel,$^1$  Pierre Illien$^1$

    $ \ $

	$^1$\textit{Sorbonne Universit\'e, CNRS, Physicochimie des \'Electrolytes et\\ Nanosyst\`emes Interfaciaux (PHENIX), Paris, France}
    
    $^2$\textit{Max Planck Institute for Dynamics and Self-Organization, Am Fa{\ss}berg 17, 37077 G\"ottingen, Germany}

\end{center}

\tableofcontents

\section{Particle-based simulations} 
\label{supp_particle_simulations}

The particle-based simulations were carried out using the LAMMPS computational package~\cite{Thompson2022}. To perform overdamped Langevin dynamics, we start from the evolution equations for the positions of the particles, which read:
\begin{equation}
   \frac{\dd\rr_n}{\dd t} = - \frac{D_0}{\kB T} \sum_{m\neq n} \nabla  U_{S_n, S_m}  (|\rr_n-\rr_m|) + \sqrt{2D_0 }  \boldsymbol{\eta}_n(t),
\end{equation}
where $D_0$ is the bare diffusion coefficient of the particles, $S_n \in \{A,B,C\}$ is the type of particle $n$, and $\boldsymbol{\eta}_n(t)$ a Gaussian white noise, of zero average, and of variance $\langle \eta_{n,i}(t)\eta_{m,j}(t') \rangle = \delta_{ij} \delta(t-t')$. Depending on the species of the particles, the potential $U_{S_n,S_m}$ is either a truncated and shifted Lennard-Jones potential:
\begin{equation}
    U_{B,B}(r) = [U_\varepsilon(r)-U_\varepsilon(r_c)] \theta(r_c-r) \qquad \text{where} \qquad U_\varepsilon(r) = 4 \varepsilon\left[\left(\frac{\sigma}{r}\right)^{12} -\left(\frac{\sigma}{r}\right)^{6} \right],
\end{equation}
and where the cutoff length is $r_c = 2.5\sigma$, or a Weeks-Chandler-Andersen potential:
\begin{equation}
    U_{A,\{ A,B,C \}}(r) = U_{C,\{ A,B,C \}}(r) = [U_{\varepsilon'}(r) +\varepsilon'] \theta(2^{1/6} \sigma-r) .
\end{equation}
The energy parameters are $\varepsilon=2\kB T$ and $\varepsilon' = \kB T$.

We use the command ‘fix brownian’ of the LAMMPS package, that integrates the evolution equations of the positions of particles thanks to a forward Euler-Maruyama scheme~\cite{Allen1987}:
\begin{equation}
   \rr_n(t+\delta t) =\rr_n(t) + \sqrt{2D_0 \delta t}  \boldsymbol{\zeta} -\delta t\frac{D_0}{\kB T} \sum_{m\neq n}\nabla U_{S_n, S_m}  (r_{mn}).
    \label{overdampedLangevin_disc}
\end{equation}
where  $\delta t= 2\cdot 10^{-4}$ and  $\boldsymbol{\zeta}$ is a random vector drawn from a normal distribution of mean $0$ and unit variance. These simulations were performed using a cubic box of length $27.144$ with periodic boundary conditions. The initial number of each species is  $N_A^0=200$, $N_B^0=800$ and $N_C^0\in \{0,900,2820,6640,10460,15280\}$. In the initial configurations, the particles are initially located on a face-centered cubic lattice. The initial configurations are equilibrated for $10^5$ timesteps.  For each set of parameters, we carried out 10 different runs with different initial conditions, and different seeds in the Brownian dynamics command ‘fix brownian’.

\section{Effect of crowding at equilibrium}
\label{supp_equilibrium}

\begin{figure}
    \centering
    \includegraphics[width=0.7\columnwidth]{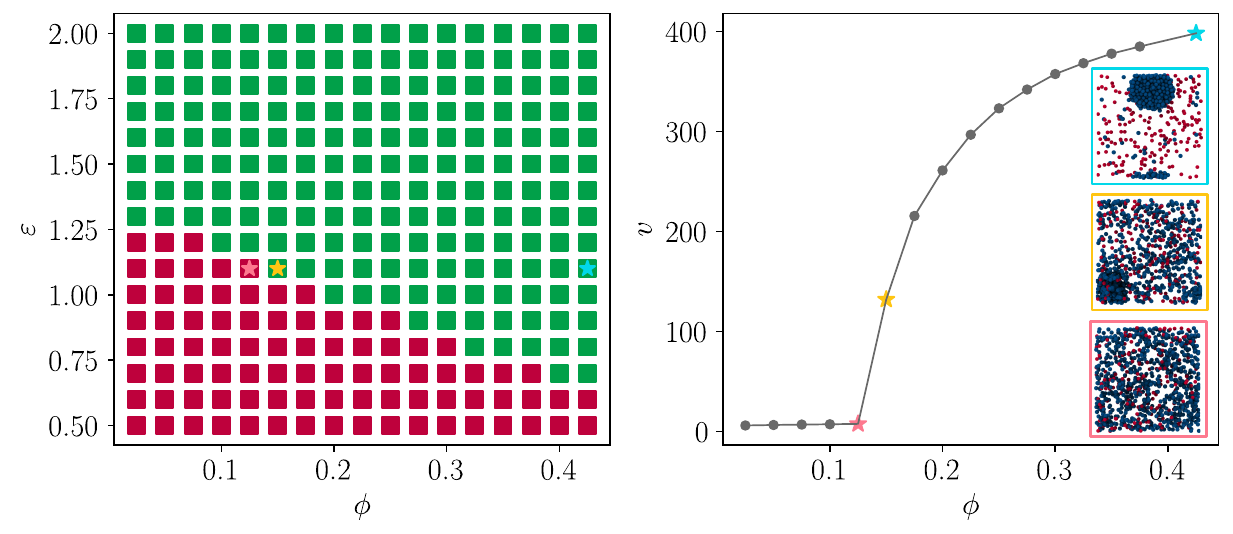}
    \caption{Left: Phase diagram for the equilibrium simulations: green squares mean that phase separation is observed in the stationary state, whereas red squares mean that no phase separation is observed. Right: Volume of the single droplet observed in the stationary state, as a function of the overall volume fraction $\phi$ for a fixed value of $\varepsilon = 1.1$. The pink, yellow and blue stars correspond to three different situations (see left panel for the ($\varepsilon$,~$\phi$)  combination). Insets are snapshots of these three different simulations.}
    \label{fig_phase_diagram}
\end{figure}

We see that the critical value of $\varepsilon$ above which phase separation is observed decreases with the overall volume fraction. Furthermore, the average droplet volume, which also corresponds to the volume of the dense phase at equilibrium (since there is only one droplet) is a increasing function of $\phi$ (Fig.~\ref{fig_phase_diagram}, right panel). This means that crowding facilitates the  phase separation in this equilibrium system. This is consistent with the expectation that crowders increase the effective attraction between $B$ particles because of depletion effects.

Concerning the impact of crowding on the mobility of the particles, we do not expect it to play any role in an equilibrium system. Indeed, at equilibrium, any change in the dynamics of the particles will not affect the stationary state, but only the time at which this state is reached from an non-equilibrated initial condition. For example, the phase diagram shown on Fig. \ref{fig_phase_diagram} is independent of the diffusion coefficients of the particles. On the contrary, for a non-equilibrium system, the steady state can be strongly influenced by the dynamics, and this is what we investigate in the main text.

\section{Phase portraits from explicit simulations}
\label{supp_phase_portraits}

\begin{figure}
    \centering
    \includegraphics[width=0.8\columnwidth]{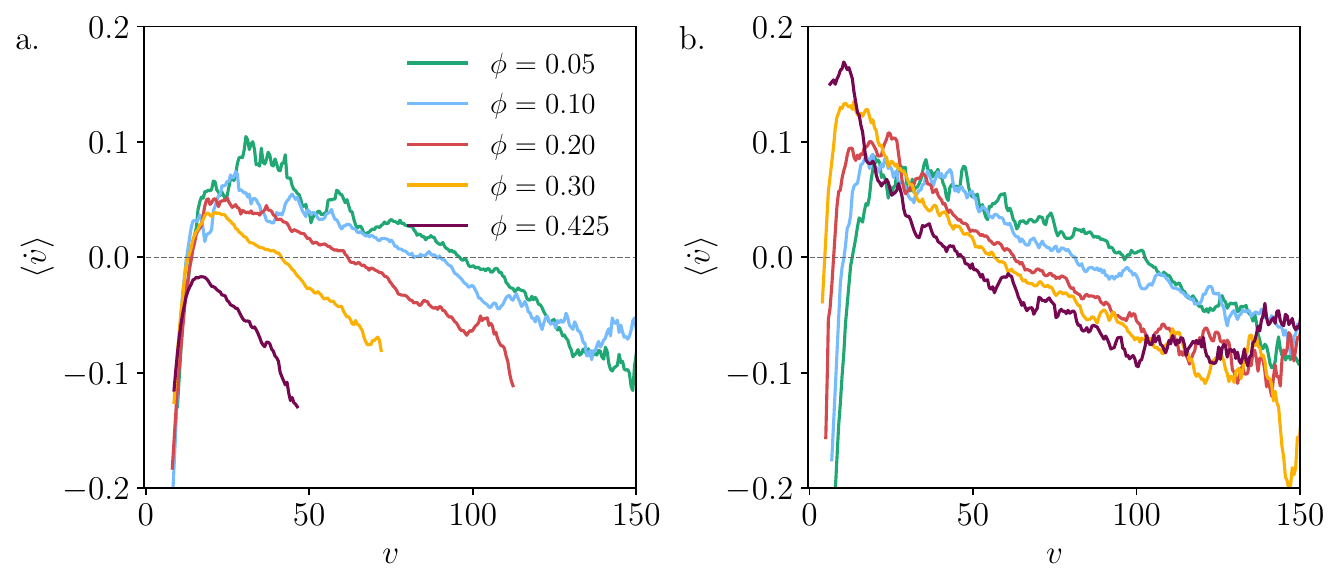}
    \caption{Phase portrait of the droplet volume, for different values of the total volume fraction, both from implicit (a) and explicit (b) particle-based simulations.}
    \label{fig_supp_phase_portrait}
\end{figure}

We show on Fig. \ref{fig_supp_phase_portrait} the phase portrait associated with the dynamics of the droplets, i.e. the typical derivative of the volume as a function of the volume. This curve typically  crosses the horizontal axis twice, therefore defining two fixed points: a stable one, that corresponds to the stable droplet radius that results from interrupted Ostwald ripening, and an unstable one, that is the critical nucleation volume $v_\text{crit}$ \cite{Berthin2025}, which is plotted as function of $\phi$, both for explicit and implicit simulations, in the main text [Fig. \ref{fig2}(d)].

\section{Field-based simulations}
\label{supp_field_simulations}

\begin{figure}
    \centering
    \includegraphics[width=0.4\columnwidth]{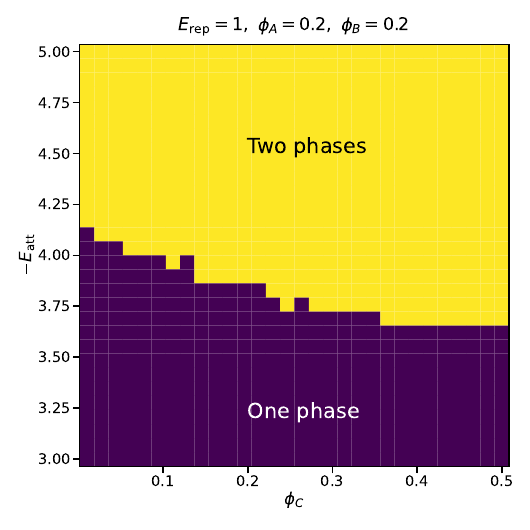}
    \caption{Phase diagram for mean field theory in the absence of reactions.}
    \label{fig_field_phase_diagram}
\end{figure}

Before examining the effects of crowding on active reactions, we first focus on the equilibrium system without reactions. To identify the equilibrium states, we minimize the total free energy using a recently developed package \cite{florypackage}, which determines whether the system forms a single phase or undergoes phase separation for a given set of parameters. To find conditions that reproduce behavior similar to particle-based simulations, we vary the interaction strength $E_\text{att}$ and the crowder volume fraction $\phi_C$. The resulting equilibrium phase diagram, shown in Fig. \ref{fig_field_phase_diagram}, reveals that a weaker attraction (lower $-E_\text{att}$) is required to induce phase separation as $\phi_C$ increases. This trend is qualitatively consistent with the particle-based simulations results shown in Fig. \ref{fig_phase_diagram}. 

We perform numerical simulations of the reaction-diffusion equation:
\begin{equation}
    \partial_t \phi_i = \nabla \cdot (\lambda_i \phi \nabla \mu_i) + R_i
\end{equation}
on an equidistantly discretized grid, using second-order finite-differences to approximate differential operators \cite{zwicker2020py}. The chemical potential gradients $\nabla \mu_i$ are evaluated on a staggered grid to ensure material conservation. Time integration is carried out using an explicit Euler scheme. The simulation domain has a size $L=150 \kappa$ with $300$ grid points, and the system is evolved up to $t=100000$. We define $\langle \xi \rangle$ as the total area where $\phi_B > (\phi_B^{\max} + \phi_B^{\min}) / 2$, which we interpret as the interior of the droplets. Here, $\phi_B^{\max}$ and $\phi_B^{\min}$ denote the maximum and minimum values of $\phi_B$, respectively. 

\begin{figure}
    \centering
    \includegraphics[width=0.8\columnwidth]{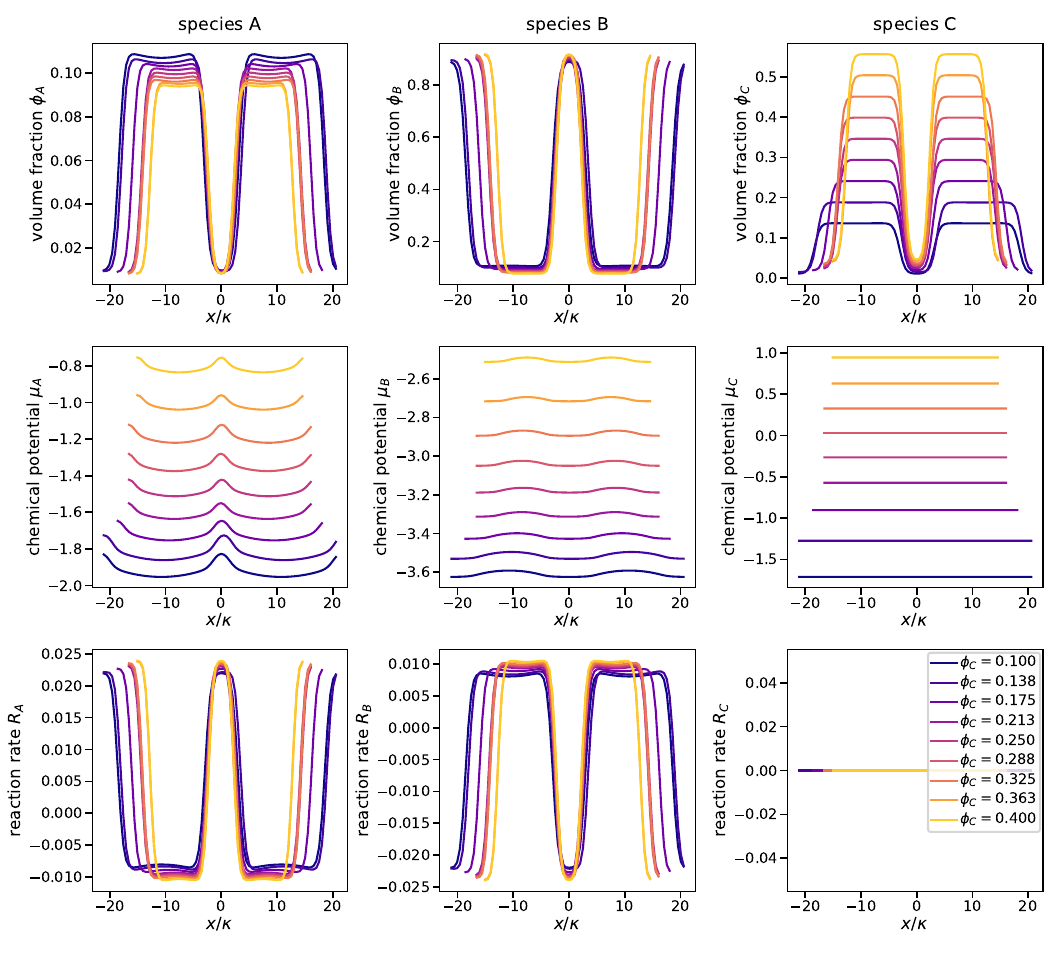}
    \caption{Profiles of volume fractions (first row), chemical potentials (second row), and total reaction rates (last row). The parameters are the same as in Fig. \ref{fig4} in the main text, and the reaction rates are the ones given in Eq. \eqref{eq:R} of the main text. For clarity, we plot two periods and shift the center of droplets to $x=0$. }
    \label{figure_profiles}
\end{figure}

Figure \ref{figure_profiles} shows the final profiles over two pattern periods, including volume fractions, chemical potentials, and reaction rates for various values of $\phi_C$. 

\begin{figure}
    \centering
    \includegraphics[width=0.7\columnwidth]{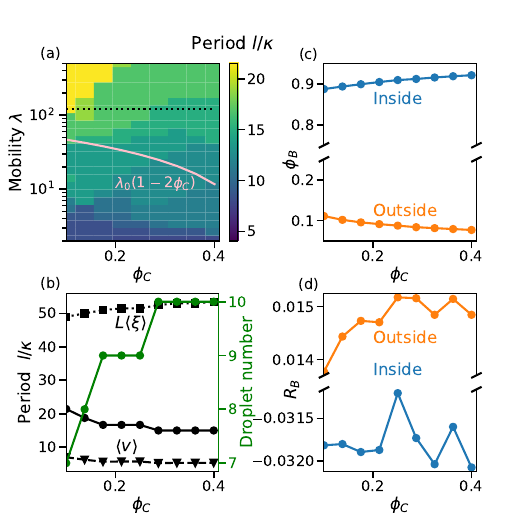}
\caption{(a) The pattern period in the field-based simulation using reaction Eq.~\eqref{eq:R_Heaviside} , shown as a color gradient, as a function of the density of crowders $\phi_C$ and of the mobility $\lambda$ of species $A$, $B$ and $C$, which are assumed to be equal and which are tuned independently of $\phi_C$. The pink solid line shows the typical dependency of the mobility over $\phi_C$ as implemented in the implicit particle-based simulations. (b) For a fixed mobility $\lambda=\lambda_0=58.6$ (indicated by the black dotted line in panel (a)), the pattern period and the number of droplets are plotted as functions of the crowder density $\phi_C$ in black and green, respectively. 
(c) The maximum value of $\phi_B$ inside droplets at the droplet center $x_1$ and the minimum value of $\phi_B$ outside droplets  at the midpoint between two droplets $x_2$. (d) The net reaction rate $R_B$ at $x_1$ and $x_2$ in orange and blue, respectively. Additional parameters are $w_B=-4$, $\phi_A+\phi_B=0.4$, $k_a=k_p=0.01$, $\Delta\mu=4$, and $\kappa=1$.}
    \label{fig_field_phase_diagram_theta}
\end{figure}

\begin{figure}
    \centering
    \includegraphics[width=0.7\columnwidth]{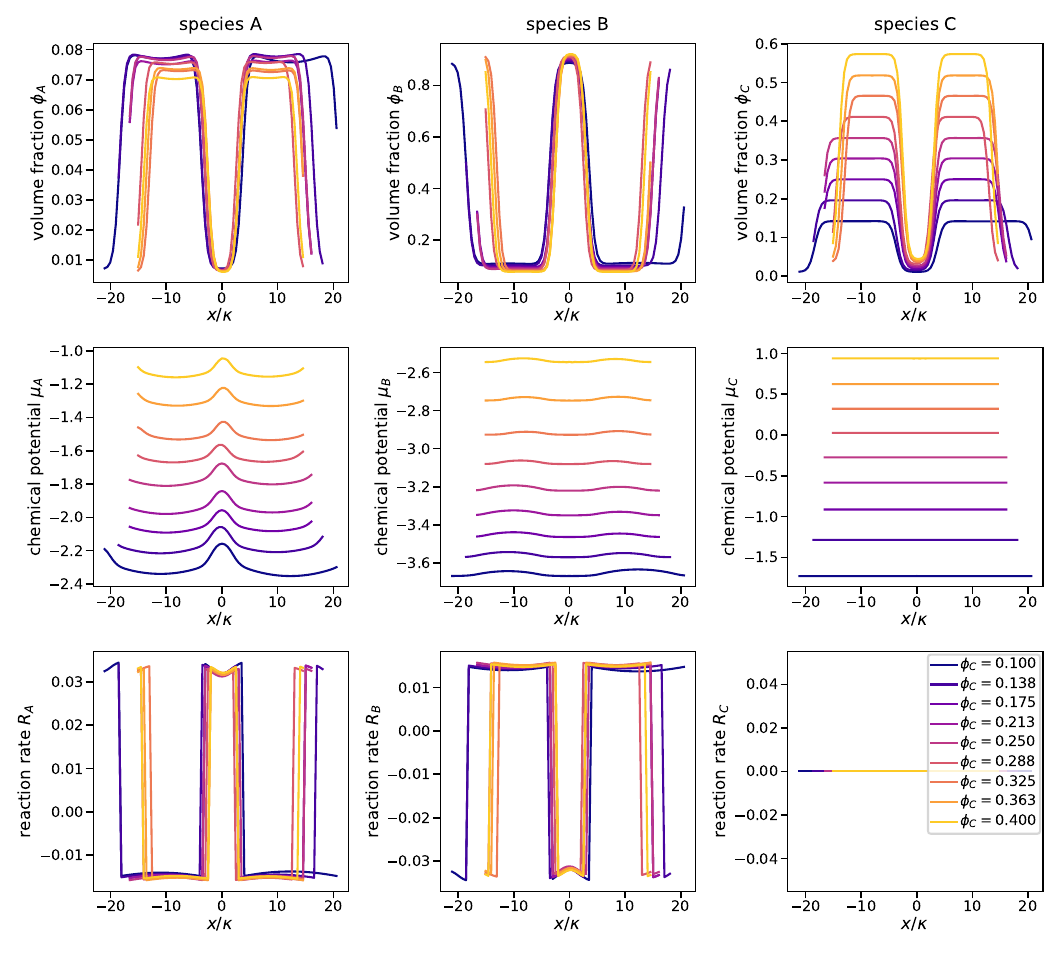}
    \caption{Profiles of volume fractions (first row), chemical potentials (second row), and total reaction rates (last row) using reaction rates given by Eq.~\eqref{eq:R_Heaviside} . The parameters are the same as in Fig. S3. For clarity, we plot two periods and shift the center of droplets to $x=0$. }
    \label{figure_profiles_theta}
\end{figure}

For completeness, we also replace the reaction by 
\begin{align}
    \label{eq:R_Heaviside}
    R_B=-R_A=2k_p\,\theta\Bigl(\frac12-\phi_B\Bigr)\sinh \Bigl(\frac\beta2(\mu_A-\mu_B)/2\Bigr)
    +2k_a \,\theta\Bigl(\phi_B-\frac12\Bigr)  \sinh\Bigl(\frac\beta2(\mu_A-\mu_B-\Delta\mu)\Bigr),
\end{align}
where $\theta(x)$ is the Heaviside function, and show the effects of crowders in Fig. \ref{fig_field_phase_diagram_theta} (see also Fig. \ref{figure_profiles_theta} for the associated profiles of volume fractions, chemical potentials, and reaction rates). The overall trend is consistent with Fig. 2 in the main text, suggesting that the model is robust to changes of the precise functional form of the prefactors of the $\sinh$ functions.

%